# THE SUMMARY OF EXPERIMENTAL RESULTS ON THE OBSERVATION OF A GAMMA RESONANCE OF THE LONG-LIVED ISOMER Ag-109m

A.V. Davydov

*State Scientific Centre of Russian Federation A.I. Alikhanov Institute of*

*Theoretical and Experimental Physics, Moscow*

andrey.davydov@itep.ru

The data obtained in 11 experiments performed up to now with gamma sources made of silver metal doped by $^{109}$Cd show that there is no large broadening of $^{109m}$Ag Mössbauer gamma line with energy of 88.o3 keV, that is the theoretically predicted gamma line broadening by ~ $10^5$ times as compared with a natural width via the dipole-dipole interaction of nuclear magnetic moments is absent. The designed in ITEP instrument of quite new type – so called "gravitational gamma spectrometer" permitted to observe the form of $^{109m}$Ag gamma resonance which turned out to be ~ $10^8$ times narrower than that of well known nuclide $^{57}$Fe. Some ideas are discussed as an attempt to explain this situation.

**1. Preface.**

Both silver isotopes, $^{107}$Ag and $^{109}$Ag, have the similar nuclear properties. In particular these isotopes have the long-living isomeric states with mean life times about 1 min that is with natural width about $10^8$ times narrower than that of the 14.4 keV state of the well known Mössbauer nuclide $^{57}$Fe. Terrestrial gravity shifts the energies of corresponding gamma resonances by their natural widths if the vertical positions of gamma emitting and absorbing nuclei differ by 1 micrometer. The Earth magnetic field splits the gamma-lines of silver isomers into 14 components and the intervals between them exceed the natural gamma line width by ~ $10^6$ times. The low energies of gamma rays of these isomers (93.1 keV of $^{107m}$Ag and 88.03 of $^{109m}$Ag) permits to perform the Mössbauer experiments with them at the temperature of liquid helium. In accordance with existing theoretical conception the Mössbauer gamma lines of Ag isomers must be strongly broadened by the mutual interactions of the magnetic moments of neighbouring nuclei and by the interactions of conduction electron magnetic moments and those of Ag nuclei (so called dipole-dipole interactions) . The energies of these interactions change rather rapidly by energy and a sign. The maximum energy achieves the value of ~ $10^{-12}$ eV.

The first experiments on the gamma resonant excitation of $^{107m}$Ag and $^{109m}$Ag isomeric states were performed by our ITEP group in sixties-seventies of past century (see [1]). Their results did not contradict to the theoretical prediction about large ( ~ $10^5$ times) broadening of corresponding

Mössbauer gamma lines caused by dipole-dipole interaction of nuclear magnetic moments. These experiments consisted in the irradiation of initially inactive silver absorbers by gamma rays of strong (2-3 Curies) sources at liquid helium temperature and subsequent detection of gamma activity of absorbers under the conditions of low background. In the case of $^{107m}$Ag the sources were fabricated by cyclotron irradiation of silver targets by protons which caused the reaction $^{107}$Ag(p,n)$^{107}$Cd. The decay of parent nuclide $^{107}$Cd with $T_{1/2}$ = 6.49 h lead to the origin of $^{109}$Ag nuclei in excited isomeric state. The gamma sources for the experiments with $^{109m}$Ag were made by the reactor neutron irradiation of the diamagnetic alloy Ag-Pd samples (the absorbers for these experiments were made of the same alloy). In this case the reaction $^{108}$Pd(n,γ)$^{109}$Pd was used. The half-life of $^{109}$Pd is equal to 13.427 h. The small values of $T_{1/2}$ of both $^{107}$Cd and $^{109}$Pd did not permit to perform a long time annealing of irradiated targets. So one can not exclude that the very small effects of gamma resonant activation in these experiments were caused not only by the gamma line broadening but also by the residual isomeric shifts between the emission gamma line of strongly irradiated gamma-sources and absorption lines of the absorbers which did not undergo such influence.

## 2. Works with gamma sources of a new type.

In 1979 the paper by German group was published [2] in which the results of the experiment on the observation of $^{109m}$Ag gamma resonance by other method were described. Authors of [2] measured the temperature dependence of $^{109m}$Ag 88 keV gamma ray yield from single crystal Ag plate in which the parent nuclide $^{109}$Cd was introduced by thermal diffusion. When gamma source was cooled from room temperature to 77 K the gamma ray yield decreased via the silver contraction. During the following cooling from 77 to 4.2 K the contraction must continue but in smaller degree because the coefficient of linear expansion decreases with decrease of the temperature. However the experiment showed that the yield decrease at second step of temperature reduction was larger than that of the first step. The surplus gamma ray absorption was ascribed to the Mössbauer resonant absorption which cross section for $^{109m}$Ag in silver metal increased by ~ 50 times at the transition from 77 to 4.2 K. The magnitude of surplus absorption permitted to estimate the broadening factor of Mössbauer gamma line which turned out to be equal to 30 (the error was not indicated). Unfortunately no explanation was given in this paper of the absence of a large gamma line dipole-dipole broadening.

Some years later the analogous experiments were performed by USA group [3-5]. They confirmed the result of [2] and obtained in three experiments the data from which the values of broadening factor followed equal to 16, 24 and 100. These authors did not explain the absence of large dipole-dipole broadening also.

Our group in ITEP performed a series of the experiments on the observation of $^{109m}$Ag gamma resonance using not only the influence of a temperature but also that of the gravity and the direction of terrestrial magnetic field. The influence of the latter was analysed in our paper [6]. As it was above-mentioned, the Earth magnetic field splits $^{109m}$Ag gamma line into 14 components the intervals between which being by ~ $10^6$ times larger than the natural width of a gamma line. Therefore each component of the emitted gamma spectrum may be absorbed in the area of corresponding component of absorption spectrum only. This leads to the strong dependence of resonant absorption probability on the angle between the direction of magnetic field and that of gamma ray detection. On the fig. 1 the dependence is shown on this angle of a factor proportional to the resonant absorption probability. It is seen that maximum of this probability corresponds to cases when this angle is equal to zero or to 180º. The influence of gravity on the probability of resonant gamma ray absorption reveals itself in the case when the vertical positions of nuclei emitting and absorbing gamma rays are different. The shift arises then between the energy of the gamma resonance equal to

$$\Delta E = E_\gamma \frac{gH}{c^2} \qquad (1)$$

Here $E_\gamma$ – gamma ray energy, $g$ – gravitational acceleration, $H$ – difference of vertical positions of emitting and absorbing nuclei, $c$ – light velocity. In the case of $^{109m}$Ag isomer this shift is equal to the natural gamma line width Γ if $H \sim 1$ micrometer.

The sketch of our experimental set-up in its final state is shown on the fig. 2. The main part of this device is a small cryostat of flowing type. Its cooled volume is suspended on the thin-walled tube through which the cooling liquid is fed and this volume is surrounded by a screen which is cooled by the waste vapors of cooling liquid. The bottom of this volume is inclined at an angle of 45º with respect to the horizontal plane. The gamma-sources were attached to the bottom inside or from outside of cooled volume (in different experiments). The main gamma source represented a silver plate (single crystal in the most of experiments) with parent nuclide $^{109}$Cd introduced by thermal diffusion. The technology of silver gamma source fabrication is described in our paper [7]. The control gamma source made of $^{57}$Co (in the first experiment only) or $^{141}$Am was placed in front of the silver one. The narrow gamma beams from the attached sources were detected by two planar HP Ge-detectors (in the final set-up version) in the horizontal and vertical directions. The pair of Helmholz rings was mounted coaxially with a cryostat. They were used for the compensation of the vertical component of Earth magnetic field in the area of gamma sources and for corresponding creation of optimal conditions for observation of $^{109m}$Ag gamma ray resonant absorption in the horizontal gamma beam.

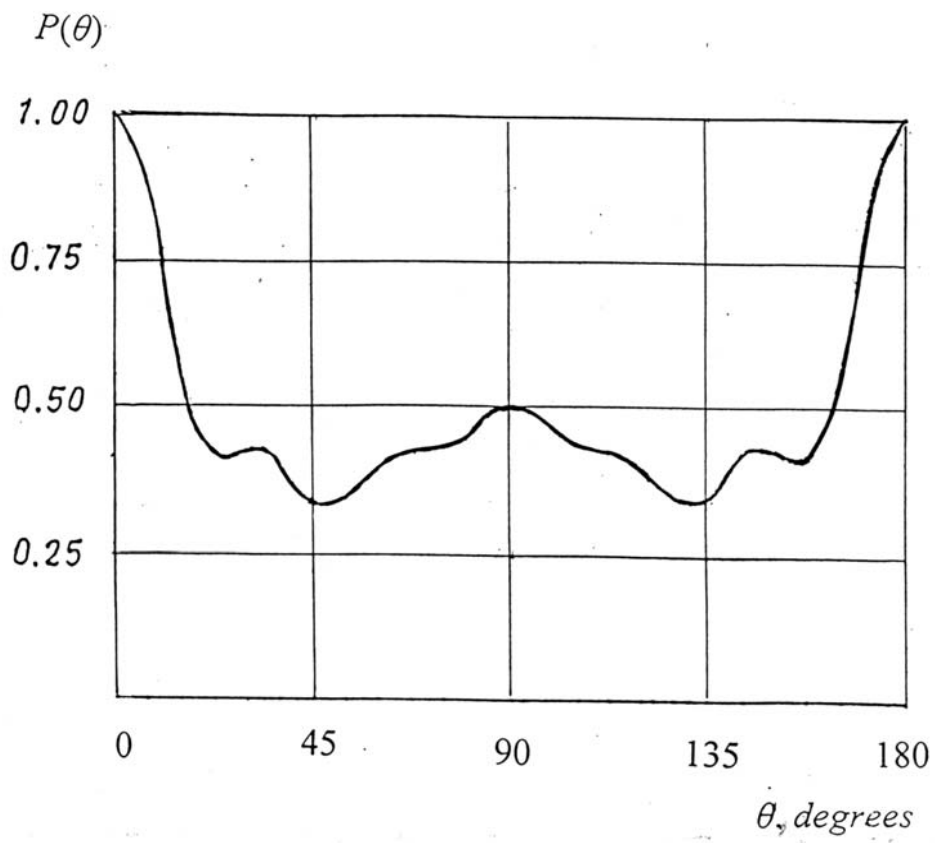

Fig. 1. Factor $P(\theta)$ proportional to the resonance absorption probability of $^{109m}$Ag gamma rays. $\theta$ – angle between the intensity vector of magnetic field and the direction in which gamma rays are detected.

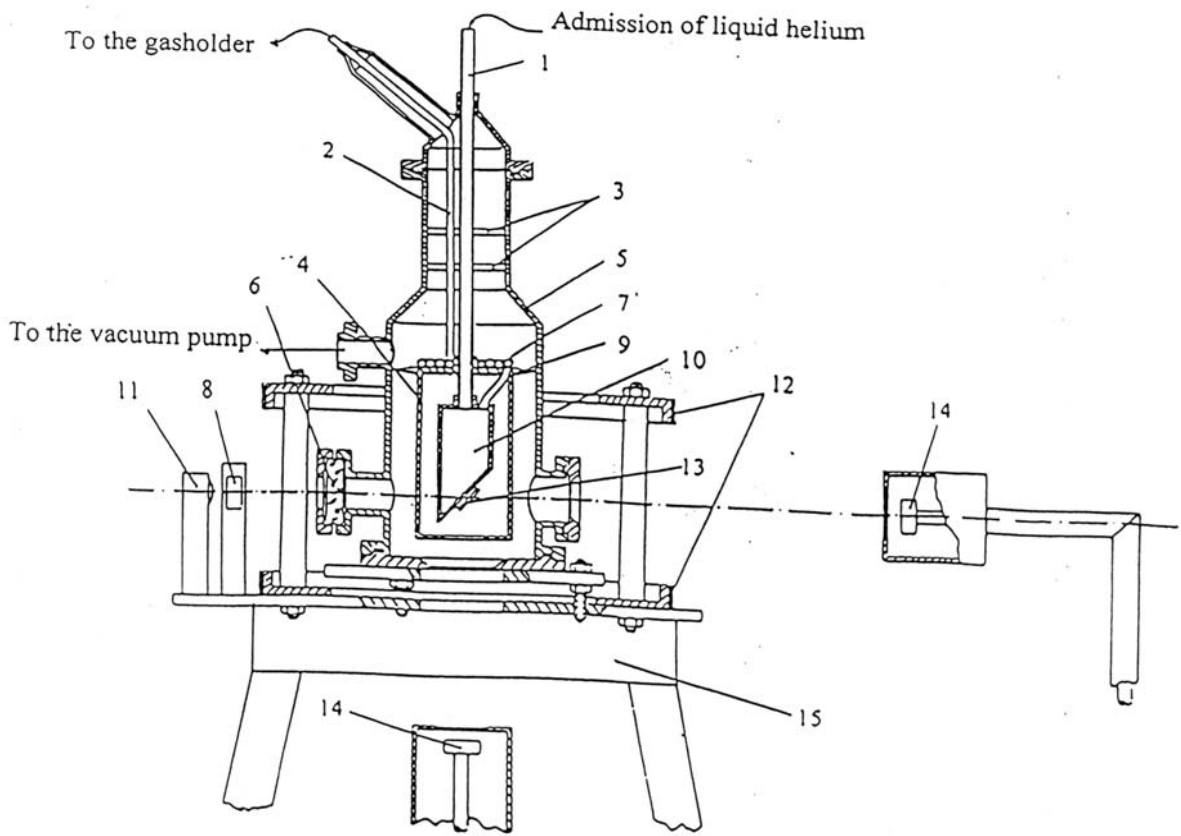

Fig. 2. Sketch of the set-up for observation the Mössbauer resonant absorption of $^{109m}$Ag gamma rays in its final version. 1 – Thin-walled tube on which the volume for liquid helium is suspended. It serves simultaneously for cryogenic liquid admission. 2 – Tube for release of helium vapor. 3 – Round quartz spacers preventing the tube bend. 4 – Anti-thermal screen of helium volume. 5 – Cryostat body. 6 – Glass window. 7 – Rolled into spiral tube for cooling of helium volume screen by evaporating helium. 8 – Mirror. 9 – Antideformation rests. 10 – Liquid helium volume. 11 – Illuminant. 12 – Helmholz rings. 13 – Gamma sources. 14 – HPGe-detectors. 15 – Support.

The glass windows in the body of the cryostat permitted to watch the possible deformational shifts of cooled volume and measure them by means of a theodolite with a precision of ~ 20 micrometers.

The first experiment [8] on this set-up was performed when it was not equipped yet by the Helmholz rings and by the gadgets for watching the inner deformation. As a main gamma source the round plate was used of single crystal high purity (99.999 %) silver 25 mm in diameter and 1 mm in thickness. The detectors were of coaxial Ge(Li) type. The measured values were the ratios $R$ of count numbers of $^{109m}$Ag gamma line (88.03 keV) and that of the control gamma source ($^{57}$Co in Cu foil, 122.1 and 136.5 keV). The temperature of gamma sources cooled by liquid helium was ~ 12K. The results of this experiment are the following:

Horizontal direction:    $$\frac{R(77\,\mathrm{K}) - R(12\,\mathrm{K})}{R(77\,\mathrm{K})} = 0.00064 \pm 0.00044$$

Vertical direction:    $$\frac{R(77\,\mathrm{K}) - R(12\,\mathrm{K})}{R(77\,\mathrm{K})} = -0.00047 \pm 0.00051$$

Of course one can not consider these data yet as the proof of the $^{109m}$Ag gamma ray resonant absorption in the horizontal gamma beam but they do not contradict such picture. The shown value for the horizontal gamma beam agrees with a result of [2[ (0.1 %) but one has to leave by three standard deviations to obtain 0.1 % as a vertical gamma beam result.

Our second experiment [9] was made with the same silver gamma source but the control source was changed by the one on the base of $^{241}$Am. It was made as a disc of chromatographic paper 25 mm in diameter impregnated with aqueous solution of americium nitrate, dried and hermetically packed into Al foil by cryostable glue. The Helmholz rings were added and it was possible to watch the deformations of inner cryostat parts. The Ge(Li) detector of the horizontal gamma beam was changed by small capsulated planar detector which was made from n-type germanium by means of strong gamma irradiation. Such detectors were being made during some time in former USSR. They could be kept at room temperature during many years without a loss of its spectroscopic properties. Unfortunately the activity of main gamma source decreased strongly towards the start of this experiment. So we could not perform the measurements switching the Helmholz rings on and off during the reasonable time and were forced to work with rings constantly switched on, that is under the optimal condition for the observation of resonant gamma ray absorption in the horizontal gamma beam. The measured values were again the ratios $R$ of count numbers corresponding to $^{109m}$Ag and $^{241}$Am gamma lines. After taking into account the corrections connected with deformations of inner cryostat part and with the contraction of cooled silver the next result was obtained for the horizontal gamma beam:

$$\frac{\Delta R}{R} = \frac{R(\text{room } T^o) - R(4.2 \text{ K})}{R(\text{room } T^o)} = 0.00300 \pm 0.00096$$

Corresponding value for the vertical gamma beam does not exceed the bounds of experimental errors.

The resonant absorption cross section for Mössbauer gamma line of $^{109m}$Ag of natural width in absence of splitting and any shift between emission and absorption lines is the following:

$$\sigma_o = \frac{\lambda^2}{2\pi} \frac{2J_e + 1}{2J_0 + 1} \frac{a}{1 + \alpha_t} f^2 = 6.35 \times 10^{-23} \text{ cm}^2 \qquad (2)$$

Here $\lambda$ – wave length of gamma radiation, $J_e$ and $J_o$ – spins of the excited and ground nuclear states correspondingly, $a$ – relative part of resonant isotope in natural mixture, $\alpha_t$ – total coefficient of inner conversion, $f$ – probability of gamma ray recoilless emission (absorption) in silver metal equal to 0.0535 at 4.2 K. If the emitting and absorbing nuclei are in the same magnetic field directed horizontally then this cross section decreased by 64/17 times in the horizontal gamma beam and by 128/17 times in the vertical one (see [6]). For gamma quanta with energy $E_\gamma$ the resonant absorption cross section by nuclei which absorption line has a Lorentzian form with a width $\Gamma = \Gamma_o k$ ($\Gamma_o$ – the natural width, $k$ – the broadening factor) is equal to

$$\sigma = \frac{\sigma_o}{k} \frac{\Gamma^2 / 4}{(E_\gamma - E_o)^2 + \Gamma^2 / 4} \qquad (3)$$

Here $E_o$ – energy corresponding to the maximum of the resonant absorption cross section The $\sigma_o$ value in (3) corresponds to the angle between the magnetic field direction and that of the gamma ray detection.

The spectrum of absorbed gamma quanta normalized to unity and also represented by Lorentzian contour with the width $\Gamma$ shifted with respect to $E_o$ by an energy $S$ is expressed as

$$N(E_\gamma) = \frac{2}{\pi \Gamma} \frac{\Gamma^2 / 4}{(E_\gamma - E_o - S)^2 + \Gamma^2 / 4} \qquad (4)$$

The averaging of the cross section (3) over the spectrum (4) yields

$$\sigma = \frac{\sigma_o}{2k} \frac{1}{(1 + S^2 / \Gamma^2)} = \frac{\sigma_o}{2k} \frac{1}{\left[1 + \left(S / \Gamma_0 k\right)^2\right]} \qquad (5)$$

The gravitational shift (1) in the case of $^{109m}$Ag gamma rays is equal to

$$S = 0.153472 \times 10^{-24} H \qquad (6)$$

Here $H$ is in cm. Since the width $\Gamma_o$ is equal to $1.16 \times 10^{-17}$ eV = $1.856 \times 10^{-29}$ erg for $^{109m}$Ag, the expression (5) can be rewritten with allowance for (6) as

$$\sigma = \frac{\sigma_o}{2k} \frac{1}{\left[1 + 0.683759 \times 10^8 \left(H/k\right)^2\right]} \quad (7)$$

Small divergence of vertical gamma beam weakly influences on the gamma ray transmission (via very small increase of mean length of the gamma-ray way through a silver only) and one can neglect it supposing that all photons fly parallel with the vertical direction. To determine the amount of gamma quanta reached the detector one has to know first of all the distribution of atoms of parent nuclide $^{109}$Cd over the gamma source thickness. If the Cd diffusion goes equally from both sides of silver plate then this distribution would be the following:

$$I(x) \sim e^{-rx^2} + e^{-r(d-x)^2} \quad (8)$$

Here $x$ – an axis perpendicular to the silver plane, $d$ – silver plate thickness. Parameter $r$ may be determined by the comparison of gamma ray and X-ray yields for given silver gamma source and for thin source which could be made for example by impregnation of a filter paper with $^{109}$Cd solution.

The estimation of the expected resonant absorption effect in the vertical gamma beam may be made using the following expression which is correct if the silver plate is disposed horizontally (however if one can neglect the gamma beam divergence it would be correct for inclined plate also but $d$ must be changed by $d\sqrt{2}$):

$$N_{v1} \sim \int_0^d \left\{ \left[e^{-ry^2} + e^{-r(d-y)^2}\right] e^{-\mu_e(d-y)} e^{-\int_y^d \frac{\sigma_o \nu}{2k} \frac{dy'}{\left[1 + 0.683759 \times 10^8 \left(\frac{y'-y}{k}\right)^2\right]}} \right\} dy \quad (9)$$

Here $y$ – co-ordinate of the point of gamma quantum emission. The corresponding axis is directed downward and its origin is in the centre of upper surface of silver plate, $y'$ – co-ordinate of the point of gamma quantum possible resonant absorption (note that gamma quantum moves downward from point $y$), $\nu$ – number of silver atoms in 1 cm$^3$, $\mu_e$ – coefficient of usual (non resonant) gamma ray absorption. After the integration in the last exponent and substitution of $\sigma_o$ and $\nu$ values the expression (9) receives the following form:

$$N_{v1} \sim \int_0^d \left\{ \left[e^{-by^2} + e^{-b(d-y)^2}\right] e^{-\mu_e(d-y)} e^{-2.93 \times 10^{-5} arctg\left[0.826897 \times 10^4 \left(\frac{d-y}{k}\right)\right]} \right\} dy \quad (10)$$

It is seen that resonant absorption effect is determined by the last exponent. Accepting

$d = 0.1$ cm and $y = 0.05$ cm (mean value) one obtains for the last exponent the value 0.999953. Such small difference from unity lies in the limits of our experimental errors. It is interesting that the value of resonant absorption effect in the vertical gamma beam depends very weakly on the gamma line broadening factor. In particular the value of the last exponent is equal to 0.999954 if $k = 10$. This is connected with following situation. The cross section of resonant absorption decreases inversely proportional to $k$ but the length of the way on which such absorption is possible increases proportional to $k$.

When one consider the resonant gamma ray absorption in the horizontal gamma beam it is preferably to take into account for the divergence of this beam that is to determine the angle $\beta$ for each photon way from source to detector with respect to the horizontal plane. As one can see at the fig 3, this angle is equal to

$$\beta = arctg\left[\frac{y_2 - y_1}{\sqrt{(z_2 - z_1)^2 + (x_2 - x_1)^2}}\right] \qquad (11)$$

Here the co-ordinates $x_1, y_1, z_1$ relate to the point $a$ where a gamma quantum is emitted and $x_2, y_2, z_2$ are the co-ordinates of the point $c$ where a photon hits the window of a detector. To receive the length of the straight line segment $ab$ along which photon moves inside of the silver plate one has to determine the co-ordinates of the point $b$ – the point of intersection of straight line $ac$ and a plane of a silver plate surface looking on the detector. General equation of a plane is the following:

$$Ax + By + Cz + D = 0 \qquad (12)$$

and that of the straight line passing through the points $x_1 y_1 z_1$ and $x_2 y_2 z_2$ [10] is

$$\frac{x - x_1}{x_2 - x_1} = \frac{y - y_1}{y_2 - y_1} = \frac{z - z_1}{z_2 - z_1} \quad \text{or} \qquad (13)$$

$$\frac{x - x_1}{l} = \frac{y - y_1}{m} = \frac{z - z_1}{n} \qquad (14)$$

The co-ordinates of the point $b$ in accordance with [10] are the following:

$$x_b = x_1 - l\rho$$
$$y_b = y_1 - m\rho \qquad (15)$$
$$z_b = z_1 - n\rho$$

Here $\rho = \dfrac{Ax_1 + By_1 + Cz_1 + D}{Al + Bm + Cn} \qquad (16)$

In our case equation (12) has a view:

$$y - z + \frac{d}{\sin 45^o} = 0 \qquad (17)$$

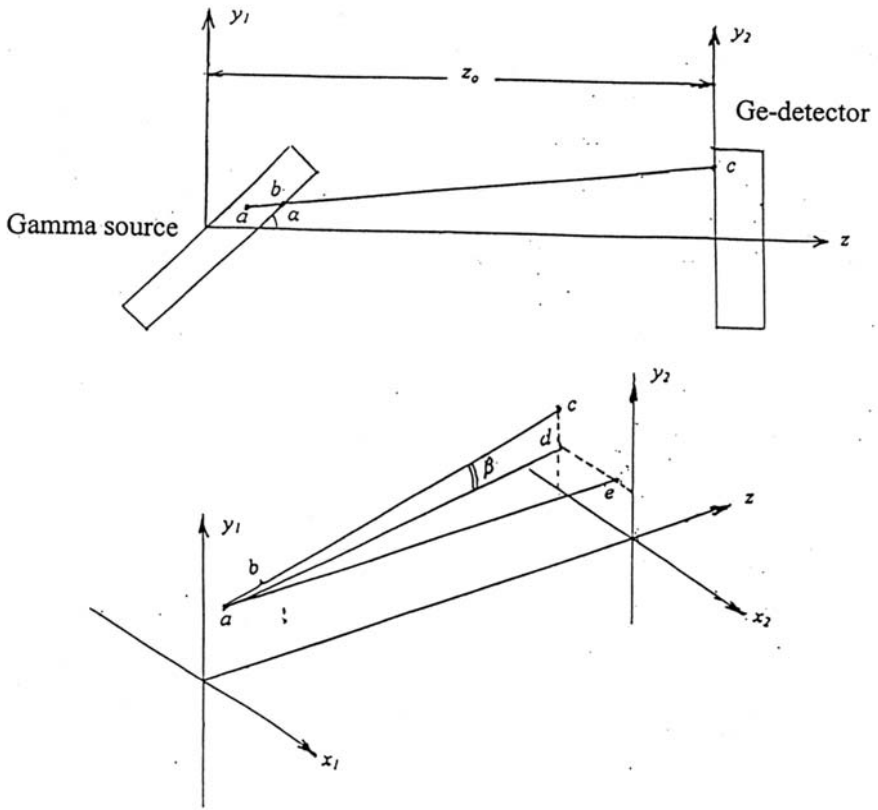

Fig. 3. Sketch elucidating the geometric quantities using at the taking into account the divergence of gamma beam.

The expressions (14-17) permit to determine the co-ordinates necessary for the further computations. The amount of $^{109m}$Ag gamma rays emitted in the volume element $dV$ of silver plate near point $a$ and hit the element $dS$ near point $c$ at liquid helium temperature may be written as

$$dN_{horiz1} = \frac{1}{4\pi L^2}\left[e^{-rp^2} + e^{-r(d-p)^2}\right]e^{-\mu_e l_{ab}} e^{-\int_0^{l_{ab}} \frac{\sigma_o v}{2k} \frac{dl}{\left[1 + 0.683759 \times 10^8 \left(\frac{l \sin\beta}{k}\right)^2\right]}} \times dVdS\cos\theta =$$

$$\frac{1}{4\pi L^2}\left[e^{-rp^2} + e^{-r(d-p)^2}\right]e^{-\mu_e l_{ab}} e^{-\frac{5.966 \times 10^{-5}}{\sin\beta} arctg\left(0.826897 \times 10^4 \frac{l_{ab} \sin\beta}{k}\right)} dVdS\cos\theta \quad (18)$$

Here $L$ – length of straight line $ac$, $p = (z_1 - y_1)\sin 45°$ (equivalent of $y$ in (10)), $l_{ab}$ – length of the segment $ab$, $\theta$ – angle between the normal to the plane of a detector window and the straight line $ac$.

The results of our second experiment are shown on fig. 4. The left scale relates to the horizontal gamma beam, the right – to the vertical one. By the shaded stripe the measured value of $\frac{\Delta R}{R}$ for the horizontal gamma beam is shown and its inaccuracy. Curve 1 is a result of $\frac{\Delta R}{R}$ calculation as a function of $k$ for this case in correspondence with (18). Curve 2 is an analogous dependence of $\frac{\Delta R}{R}$ on $k$ for the vertical gamma beam (see the right scale). It is seen that the measured and calculated values for the horizontal gamma beam are in mutual accordance at $1 < k < 3$. Note that if we did not take into account the influence of terrestrial magnetic field, we should receive for $\frac{\Delta R}{R}$ the value corresponding to $k = 35^{+19}_{-10}$ using the same experimental data. This means that our result does not contradict the data of works [2-5] which were performed without taking into account the effect of this field.

If the gamma beam is sufficiently narrow one can disregard its divergence. It is possible in this case to describe the experimental result for the horizontal gamma beam by the following expression:

$$N_{horiz.} \sim \int_0^d \left[e^{-rz^2} + e^{-(d-z)^2}\right]e^{-\mu_e(d-z)} e^{-\frac{\sigma_o v}{2k}(d-z)} dz \quad (19)$$

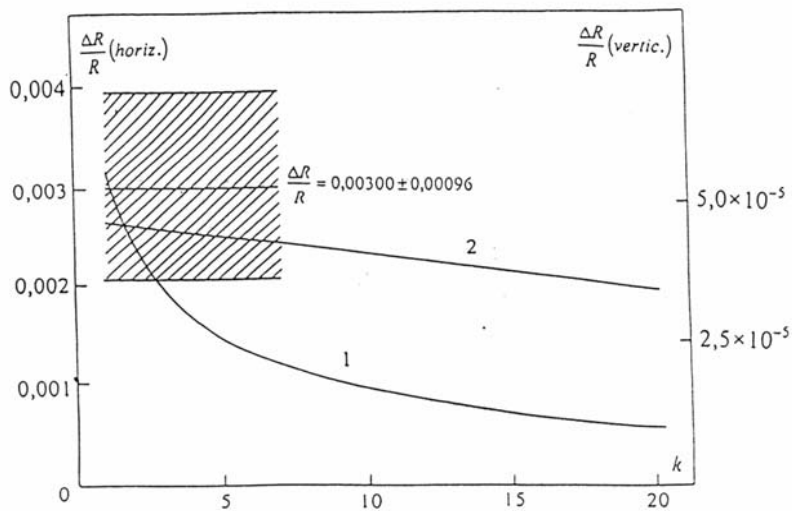

Fig. 4. The results of the experiment on the observation of $^{109m}$Ag Mössbauer gamma ray resonant absorption described in the paper [9]. Along the left ordinate axis the values of $\frac{\Delta R}{R}$ are indicated for the horizontal gamma beam (measured and calculated). The right scale corresponds to the vertical gamma beam. The solid curve 1 presents the calculated dependence of $\frac{\Delta R}{R}$ for the horizontal gamma beam on the broadening factor $k$, the curve 2 – for the vertical one. By the shaded stripe the measured value of $\frac{\Delta R}{R}$ and its inaccuracy are shown for the horizontal gamma beam. The measurements were performed at Helmholz rings constantly switched on.

The constancy of a cross section over the total way of the photon in silver is explained by the low value of relative resonant absorption effect ($\sim 10^{-3}$) and by corresponding possibility to neglect the gamma line broadening connected with selective resonant absorption.

If the measurements were performed under the conditions which excluded the possibility of observation of resonant gamma ray absorption (room or 77 K temperature) the last exponent in (19) must be removed.

The next our experiment was made under the support of INTAS [11]. In this work we for a first time used a procedure of a periodic switching of the Helmholz rings on and off in addition to the observation of a temperature dependence of gamma ray yield. New silver gamma source was used. The results were the following. The influences of the temperature decrease from 77 to 4.2 K led to the $\frac{\Delta R}{R}$ values:

Horizontal gamma beam: $\frac{\Delta R}{R} = 0.00116 \pm 0.00080$.

Vertical gamma beam: $\frac{\Delta R}{R} = 0.00062 \pm 0.00061$ (the calculated $\frac{\Delta R}{R}$ value for this gamma beam is equal to $0.00004 \pm 0.00001$ and depends very weakly on gamma line broadening).

The effects of the direction change of terrestrial magnetic field turned out to be:

Horizontal gamma beam: $\frac{\Delta R}{R} = 0.00113 \pm 0.00073$.

Vertical gamma beam: $\frac{\Delta R}{R} = 0.00006 \pm 0.00068$

The $\frac{\Delta R}{R}$ value for the horizontal gamma beam corresponds in this case to 60 % of total effect of resonant gamma ray absorption. So this effect is equal to $0.00188 \pm 0.00125$. The unification of both results for the horizontal gamma beam gives the value of relative resonant absorption effect equal to $0.00137 \pm 0.00067$.

Then we performed two experiments [12] with silver gamma sources of different types: one was made of polycrystalline silver 0.5 mm in thickness, the other – of single crystal silver. In both cases silver was of high purity (~ 99.999 %). Unfortunately we could not observe the appearance of gamma ray resonant absorption as a result of gamma source cooling. Although even at absence of resonant absorption one has to observe the decrease of gamma ray yield connected with silver contraction, we observed instead the small increase of this yield. The deformational shifts of inner cryostat part in direction of a detector were absent. We give the following explanation to this fact. The silver sources were soldered by the Cd-Bi alloy to the rather thin copper bottom of the cryostat helium volume and composed bimetallic pairs with it. The thermal expansion coefficients of silver and copper are different and therefore the deformation of very sources without shift of helium volume as a whole could be possible. However after the temperature establishment when the further deformations would be ended one could observe the influence of magnetic field direction on the gamma ray resonant absorption. The results of the experiment with a polycrystalline gamma source are shown on the fig. 5. The points under the brackets marked as $\gamma_{Ag}$ and $\gamma_{Am}$ are the ratios of the detected gamma ray intensities measured with the Helmholz rings switched on and off separately for silver and americium gamma lines. Black circles relate to the measurements at liquid helium temperatures, the open ones – to the measurements at room and 77 K temperatures. Under the brackets marked as R the ratios of these values related to Ag and Am are shown. It is seen that in the case of Ag gamma rays the effect of resonant absorption reveals in the horizontal gamma beam at 4.2 K as a decrease of R as compared with 1 and is absent over the error limits in the vertical one and at room and 77 K temperature. The data for Am gamma rays show the absence of the magnetic field direction influence in all cases in both gamma beams. The value corresponding to 60 % of the total resonant absorption of $^{109m}$Ag gamma rays determined according to the intensities of corresponding gamma line is equal to

$$(1.00007 \pm 0.00011) - (0.99959 \pm 0.00022) = 0.00048 \pm 0.00025$$

The total resonance absorption effect is equal to $0.00079 \pm 0.00041$.

On the fig. 6 the results are shown of the experiment with a single crystal gamma source. At first glance, it may seem that these results are less convincing than the data obtained for the polycrystalline silver. Indeed the error bar of the $^{109m}$Ag gamma ray intensity ratio measured at at liquid helium temperature in the horizontal gamma beam touches the line corresponding to unity. However, note that the intensity ratio of the $^{241}$Am gamma rays measured by the same detector simultaneously with the corresponding values for the $^{109m}$Ag gamma rays is significantly higher than unity. These shifts could be caused by the drift phenomena in the electronic circuit of the horizontal gamma beam detector. But on the other side the values of R show quite distinctly the presence of the resonant absorption effect which relative value is equal to

$$[(1.00011 \pm 0.00014) - (0.99943 \pm 0.00037)] / 0.6 = 0.00113 \pm 0.00067$$

These effects correspond to the following values of Mössbauer gamma line broadening factor:

$k = 22^{+25}_{-8}$ for the polycrystalline gamma source and $21^{+13}_{-6}$ for the single crystal one.

The summary result of last four described experiments is equivalent to that of one work which value of resonant gamma ray absorption effect exceeds the zero level by more than 4 standard deviations. Thus in all without exclusion nine above-mentioned experiments the data were obtained showing the absence of a large dipole-dipole broadening of Mössbauer gamma line of $^{109m}$Ag isomer. This permitted us to design and fabricate the instrument of quite new type – the gravitational gamma spectrometer for measurement the form of the gamma resonance of a long-lived isomer $^{109m}$Ag.

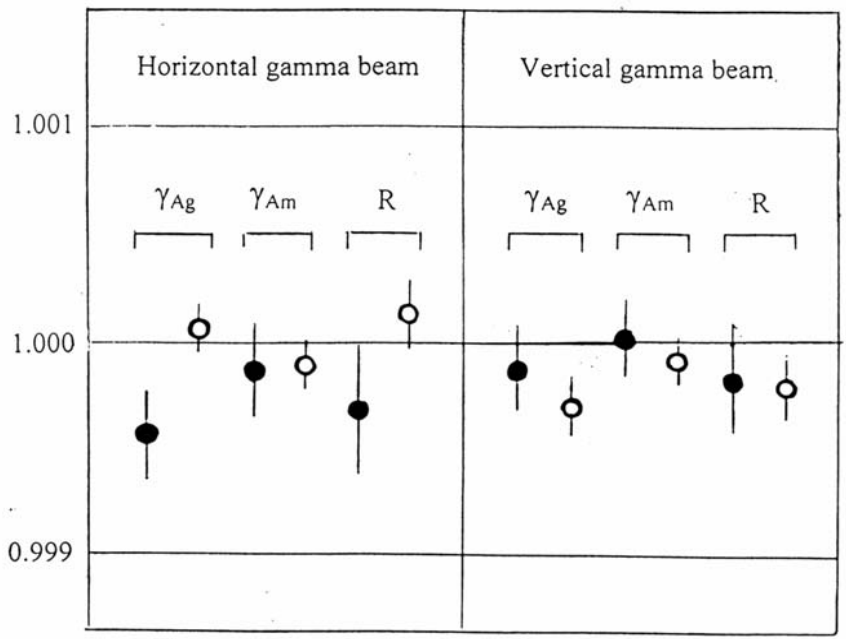

Fig. 5. The results of the experiments with polycrystalline silver gamma source and control source of $^{241}$Am performed in the work [12]. Along the ordinate axis the ratios are indicated of gamma ray intensities which were measured at Helmholz rings switched on and off. Under the brackets signed by R the ratios of quantities R are shown which are the ratios of Ag and Am gamma ray intensities ("doubled" ratios) measured simultaneously by the same detector. Open circles corresponds to the averaged results measured at room temperature and at 77 K. Black circles relate to the temperature of liquid helium.

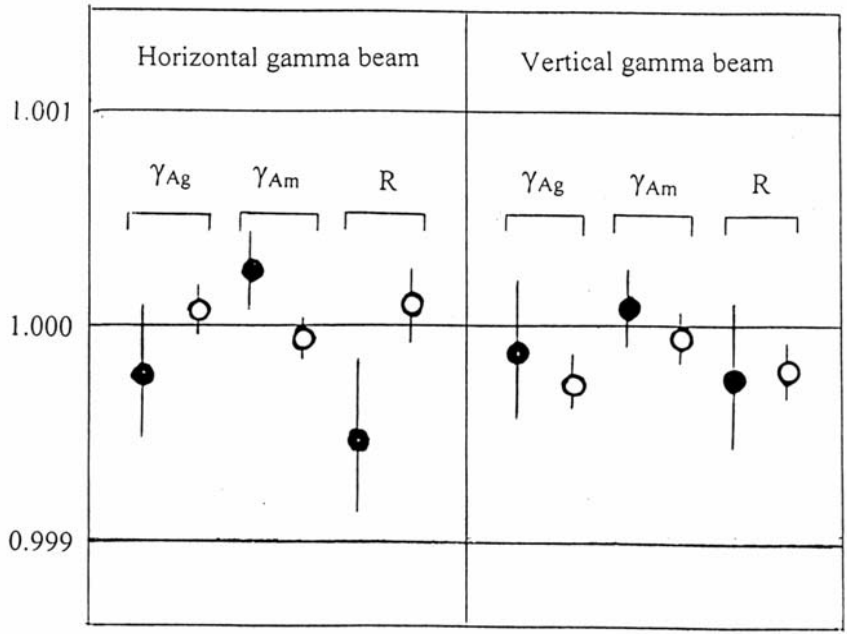

Fig. 6. The results of the experiments [12] with single crystal gamma source and control source of $^{241}$Am. All designations are the same as on the fig, 5.

**3. Start of the gravitational gamma spectrometry.**

The sketch of the first gravitational gamma spectrometer is shown on the fig. 7. On the platform which may be rotated in both directions by the angle up to 30° around the horizontal axis the cryostat and two planar HP Ge-detectors are mounted. Two Helmholz rings are mounted coaxially with a cryostat for additional revelation of gamma ray resonant absorption by changing of Earth magnetic field direction. Gamma sources of two types are placed inside of a cryostat. The main gamma source is a single crystal silver plate doped by atoms of parent nuclide $^{109}$Cd. Its dimensions are 16×24×0.74 mm$^3$. Two control $^{241}$Am gamma sources made by the above-described method are pressed to the main one from both sides of it. The detectors register the gamma rays of two energies in narrow gamma beams parallel to the plane of the platform. Experiment consists in a measurement of gamma ray intensities as a function of the rotation angle of the platform. Principle of the spectrometer work is elucidated by a fig. 8 where the cross section of silver gamma source is shown in the sloping position. When a gamma quantum emitted in the point A moves in direction of point B of its possible resonant absorption, the value $H$, the difference of vertical positions of the emission and absorption points, is gradually increases and this leads to the gradual decrease of the cross section of resonant gammaray absorption in correspondence with (7). The larger the angle of the gamma beam slope the stronger is the effective diminution of cross section. At sufficiently large angles of a slope one could not observe the effect of resonant absorption over the limits of experimental inaccuracies. If gamma beam is sufficiently narrow then one may neglect the divergence of this beam The dependence of detected $^{109m}$Ag gamma ray intensity on the slope angle $\alpha$ of the gamma beam would be the following in this case.

$$N_\gamma(\alpha) \sim \left[ e^{-rx^2} + e^{-r(d-x)^2} \right] e^{-\mu_e(d-x)} e^{-\int_x^d \frac{\sigma_o v}{2k} \frac{dl}{\left[1 + 0.683759 \times 10^8 \left(\frac{l \sin \alpha}{k}\right)\right]}} =$$

$$= \left[ e^{-rx^2} + e^{-r(d-x)^2} \right] e^{-\mu_e(d-x)} e^{-\frac{\sigma_o v}{1.653794 \times 10^4 \sin \alpha} \arctg\left[0.826897 \times 10^4 \left(\frac{d \sin \alpha}{k}\right)\right]} \quad (20)$$

The first description of the construction of a gravitational gamma spectrometer was given in our paper [13]. There were presented in particular some calculated data on the angular dependencies of detected intensities of $^{109m}$Ag gamma rays emitted from the silver plate of 1 mm thickness. It was assumed that atoms of $^{109}$Cd concentrated in thin layer of silver at the distance of 0.85 mm from the source side looking at the detector. These data are shown on the fig. 9 for several values of Mössbauer gamma line broadening factor *k*. All curves are normalized to the same magnitude at zero slope angle. However one has to remember that the cross section of resonant absorption decreases in this case inversely proportional to *k*. On the next fig. 10 the dependence is shown on *k* of the slope angle $\alpha_{1/2}$ corresponding to decrease of resonant absorption by two times. This dependence is very close to the straight line and may be presented in this case as

$$\alpha_{1/2} = 0.197\, k \tag{21}$$

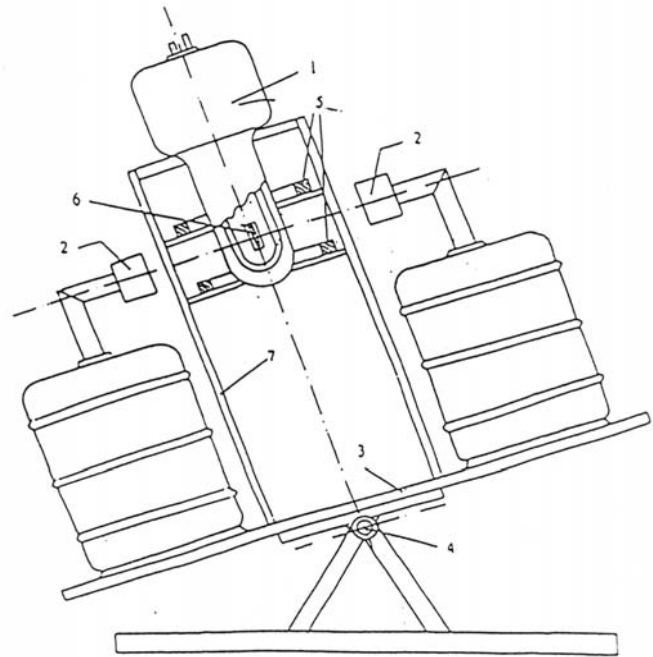

Fig. 7. Sketch of the gravitational gamma spectrometer.
  1 – cryostat, 2 – Ge-detectors, 3 – inclinable platform, 4 – axis of platform rotation,
  5 – Helmholz rings, 6 – gamma sources, 7 – support of the cryostat and Helmholz rings.

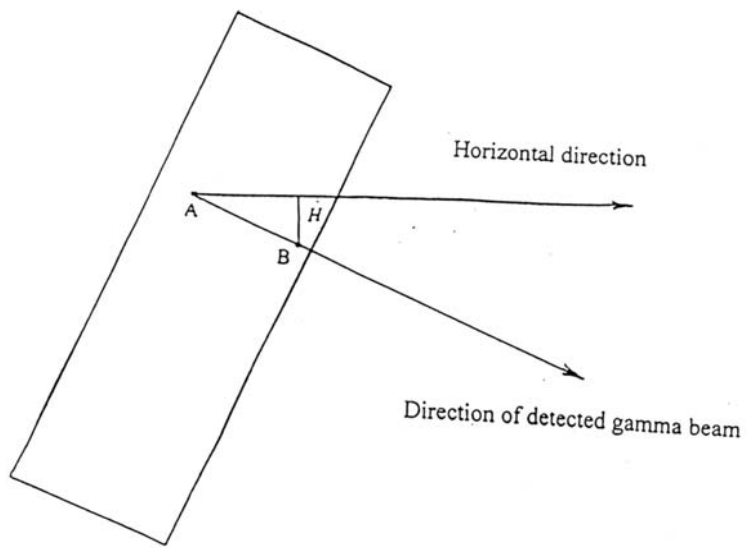

Fig. 8. Illustration elucidating the work principle of a gravitational gamma spectrometer. The cross section is shown of silver gamma source in the position when the detected gamma beam is declined from the horizontal direction. A – point of gamma quantum emission, B – point of its possible resonant absorption, $H$ – difference of vertical co-ordinates of points A and B. It grows with increase of a distance between these points.

Here $\alpha_{1/2}$ is in degrees. This expression may be used for rough estimation of *k*.

First experiment with this spectometer which was modified slightly as compared with the view presented in [13] is described in our papers [14, 15]. It consisted in the measurements of $^{109m}$Ag and $^{241}$Am gamma ray intensities as a functions of the slope angle of narrow gamma beams at liquid helium and room temperatures. Note that the gamma sources in the cryostat were directly immersed in liquid helium. The measurements were performed at each slope angle in order of B→ A → A →B where B means the measurement with Helmholz rings switched off and A corresponds to that with rings switched on. Therefore the switching was performed without the change of spectrometer position and one could not ascribe its influence to any mechanical cause. The measurements were made at the slope angles +7°, +3°, +1°, 0°, –1°, –3° and –7°. Sign + corresponds to the raising of the detector No 1 over the horizontal plane. The results of the measurements at liquid helium temperature are shown on the fig. 11 where on the part A the angular dependence of the ratio $R^+$ of Ag and Am gamma ray intensities is presented measured by detector No 1 when compensating rings were switched on. Curve 1 is calculated in correspondence with (20). It describes optimally the experimental data with corresponding value of broadening factor *k* = 11. The value of $\chi^2$ criterion is equal to 0.655 per one degree of freedom. Curve 2 is calculated for $\chi^2 = 1$. The *k* value is equal in this case to 7.7. The dotted line shows the calculated level of $R^+$ in the absence of resonant absorption. On the part B of fig. 11 the results are presented of $R^-$ measurements performed with Helmholz rings switched off. The resonant absorption of $^{109m}$Ag gamma rays must decrease in this case by 2.5 times. And really the corresponding effect is not shown over the limits of the experimental inaccuracies. On the part C the results are presented of $R^+$ measurements which were made at room temperature of gamma sources. The resonant absorption effect is obviously absent as one had to expect. Thus all expected signs of real observation of resonant gamma ray absorption are present. Unfortunately we could not observe this effect in the data obtained from detector No 2. The direction from the centre of gamma source to the centre of this detector turned out to be not parallel to the plane of the spectrometer platform. The deviation was such that only weak decrease of $R^+$ value was observed at +1° and the main part of resonant pattern was hidden in the interval between +1° and +3° where the measurements were not performed. The second experiment on the gravitational gamma spectrometer [16] was carried out without any changes of the set-up. Therefore it may be considered as a continuation of the first one. The small difference was in the number of angular positions. In addition to the slope angles enumerated above the following four positions were added: ±0.33° and ±0.67°. It was found during the measurements that americium control gamma sources were partly destroyed during more than 2 years passed from their fabrication under the influence of intensive alpha particle irradiation and possible chemical processes. The fragments of a paper impregnated by americium nitrate began to move inside their aluminium shells as result of the unavoidable pushes at switching of a platform drive on and off. This led to the unpredicted changes of corresponding counting rate. Therefore the data are presented below connected with $^{109m}$Ag gamma rays only. As regards the control role of intensity measurement of americium gamma rays, that one may be convinced in the absence of something like resonant absorption effect observing the angular dependence of americium gamma ray counting rate measured in the first experiment on the gravitational gamma spectrometer when the americium gamma sources were not damaged yet (fig. 12). These sources were not touched between two experiments. On the fig. 13 the results are shown of the measurements of the $^{109m}$Ag gamma ray intensity dependence on the slope angle by detector No 1 at the liquid helium temperature of gamma sources. At the part A the data are presented related to the regime of Helmholz rings switched on, those at the part B were measured with rings switched off. It is seen clearly the influence of the Earth magnetic field direction. The smooth curves represent the optimaldescription of experimental data. They were calculated using the expression (20). The strongly deviated point at +7° on the part B was not included in these calculations. The value of $\chi^2$ criterion for the smooth curve on part A is equal to 0.62 per one degree of freedom the broadening factor $k$ being equal to $7.0^{+7.3}_{-2.5}$. The smooth curve on part B corresponds to $\chi^2 = 0.65$ and $k = 8$ with increased errors. The dotted lines show as earlier the calculated counting rate levels in absence of resonant gamma ray absorption. The data of fig. 13A corresponds to the resonant absorption effect which differs from zero level by more than 8 standard deviations. On the fig. 14 the results are shown of the measurements performed at room temperature of gamma sources. Parts A and B relate as earlier to regimes connected with Helmholz ringsswitched on and off correspondingly. It is clearly seen that in both cases there are no decreases of counting rate near zero slope angle. Therefore both experiments with gravitational gamma spectrometer have shown that the large dipole-dipole broadening of $^{109m}$Ag gamma line is really absent. More of that the data of these experiments permit to estimate the form of $^{109m}$Ag gamma resonance. Its measured width gives a possibility to state that the resolving power is achieved which surpasses by ~$10^8$ times the resolution of existing Mossbauer spectrometers working with $^{57}$Fe nuclide.

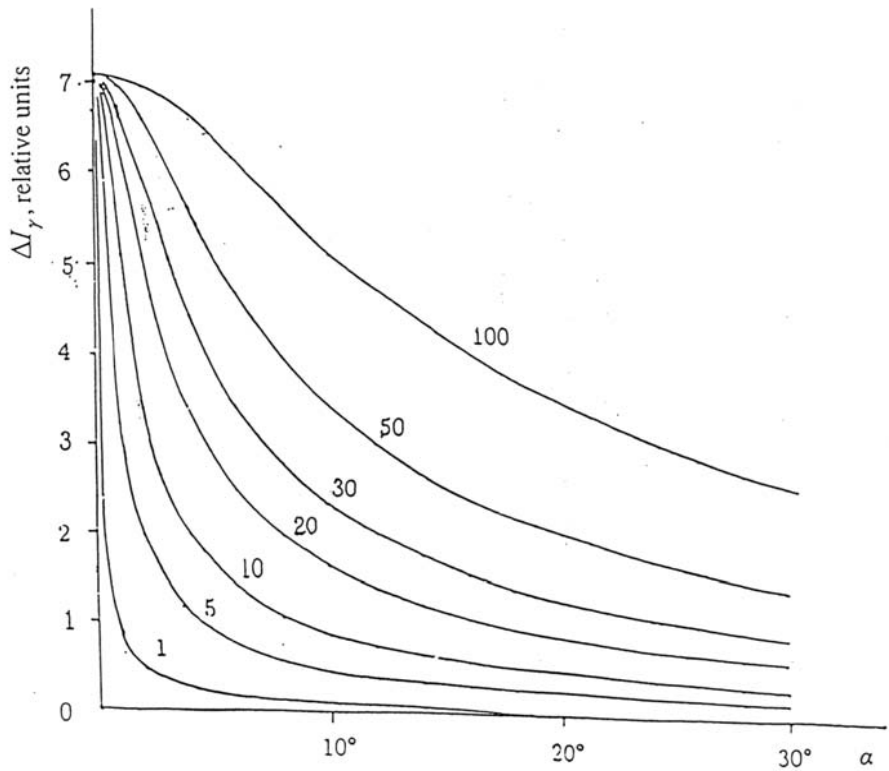

Fig. 9. The dependencies of a diminution of $^{109m}$Ag gamma ray intensity $\Delta I_\gamma$ caused by resonant self-absorption in silver on the gamma beam slope angle $\alpha$ for several values of broadening factor $k$ of Mössbauer gamma line. These values are indicated near the corresponding curves which are normalized on the same magnitude at $\alpha = 0$. However one has to remember that the real values of $\Delta I_\gamma$ are inversely proportional to $k$.

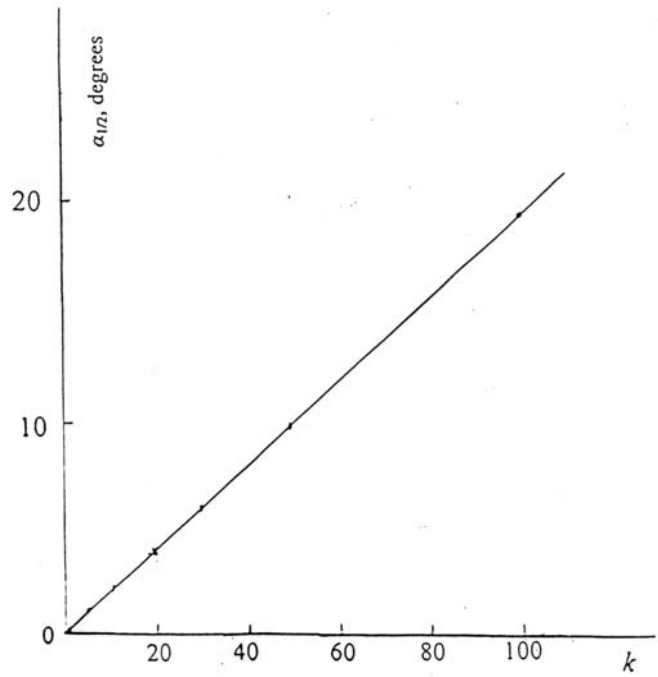

Fig. 10. The dependence of the slope angle $\alpha_{1/2}$ which corresponds to twice decrease of resonant gamma ray absorption on the broadening factor of Mössbauer gamma line $k$.

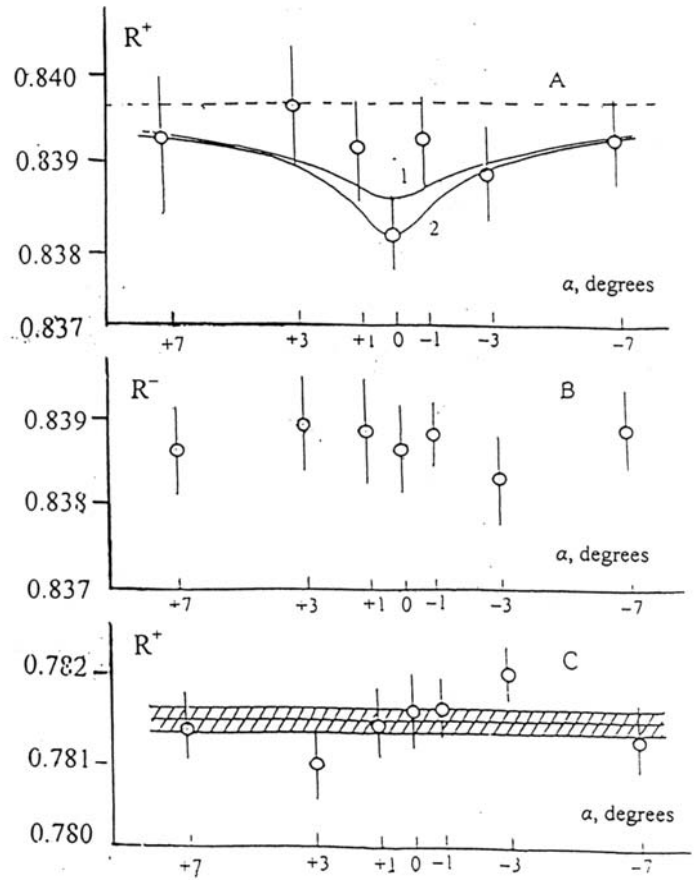

Fig. 11. The results of the first experiment performed on gravitational gamma spectrometer.
A – the dependence of the ratio $R^+$ of detector N1 counts related to $^{109m}$Ag and $^{241}$Am gamma lines on the gamma beam slope angle at the Helmholz rings switched on. Solid line 1 optimally describes this dependence. It is calculated by formula (20) and corresponds to $k = 11$ and to the value of criterion $\chi^2$ equal to 0.655 per one degree of freedom. The curve 2 is calculated for $\chi^2 = 1$ and it corresponds to $k = 7.7$. The dotted straight line shows the calculated value of $R^+$ for the case of resonant absorption absence. B – analogous dependence for the ratio $R^-$ measured at rings switched off. C – the results of $R^+$ measurement at room temperature of gamma sources. The shaded stripe shows the mean value of counts over all angular positions and its inaccuracy.
23

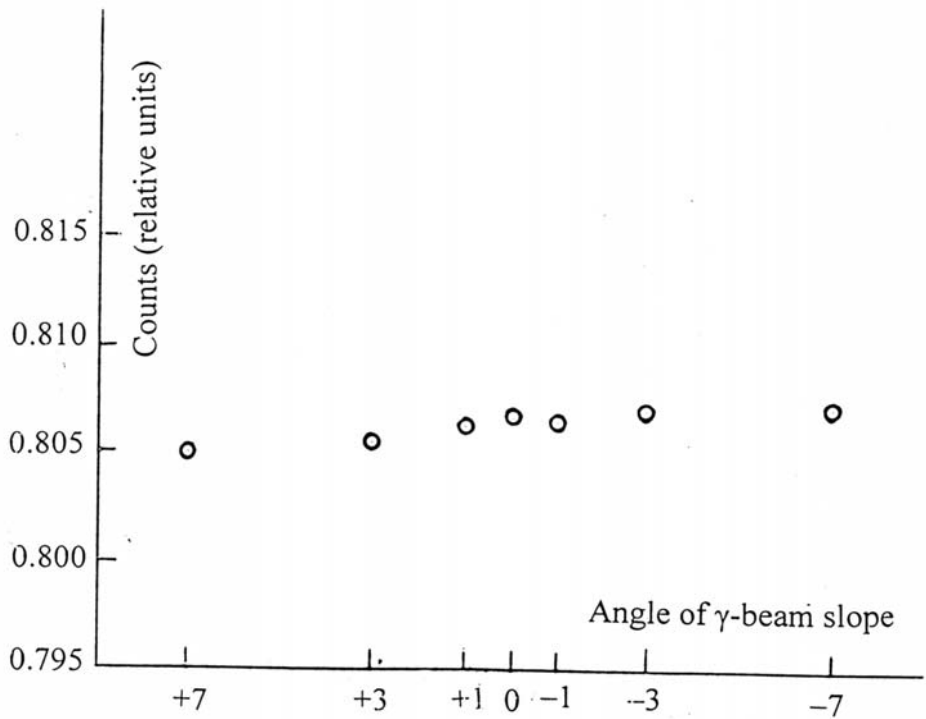

Fig. 12. The results of the measurements of the $^{241}$Am gamma ray intensity dependence on the gamma beam slope angle obtained in the first experiment with gravitational gamma spectrometer. The statistical standard deviations are equal to the circle radii.

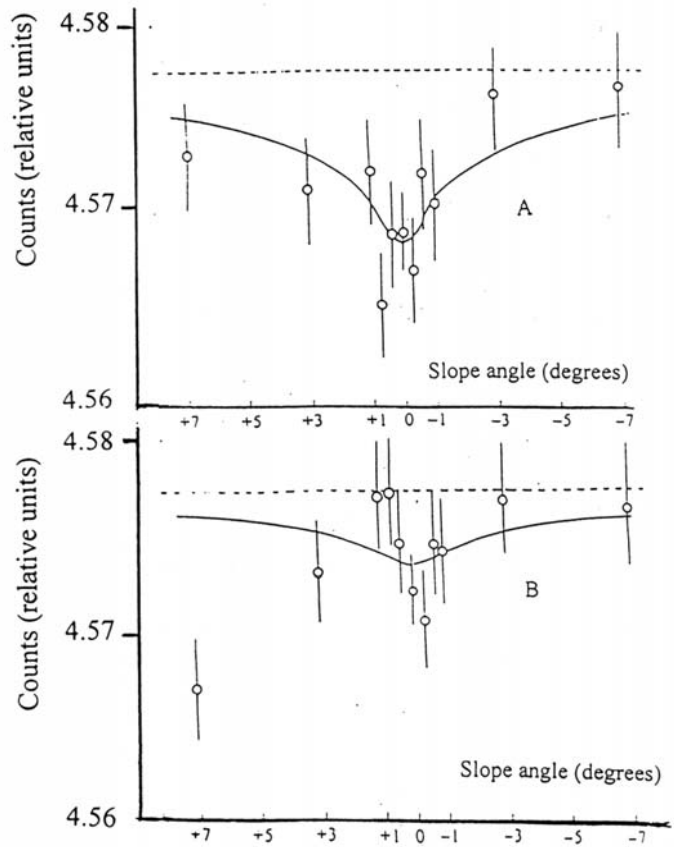

Fig. 13. The measurement results of the $^{109m}$Ag gamma ray intensity dependence on the gamma beam slope angle at liquid helium temperature obtained in the second experiment with gravitational gamma spectrometer. A – Helmholz rings are switched on. B – rings are switched off.

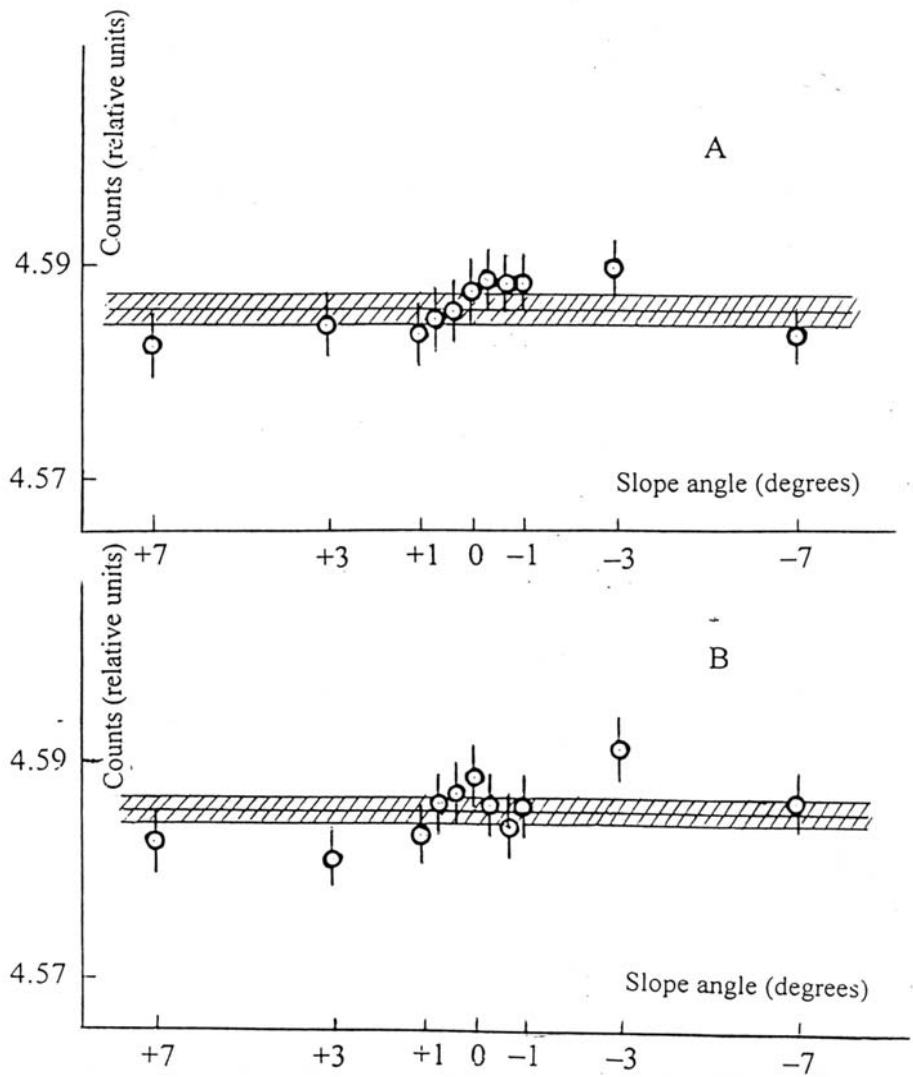

Fig. 14. The measurement results of the $^{109m}$Ag gamma ray intensity dependence on the gamma beam slope angle at room temperature obtained in the second experiment with gravitational gamma spectrometer. A – Helmholz rings are switched on. B – rings are switched off. The shaded stripes presents mean values of counts and their inaccuracies.

## 4. Conclusions

In all 11 experiments without exclusion performed until now with gamma sources fabricated by thermal diffusion doping of silver by $^{109}$Cd the data were obtained which have shown with different accuracy the presence of resonant gamma ray self absorption at the temperature of 4.2 K and near it. This means in its turn that the large dipole-dipole broadening of $^{109m}$Ag Mössbauer gamma line is absent. This situation is possible, independently of yet unknown reasons, on condition that gamma quantum is being emitted during the time much larger than the characteristic time $\tau_{dd}$ of a change of dipole-dipole interaction energy, and most probably during the total life time of the nucleus in excited state (see also [17, 18] about it).

A.B. Migdal states in his book [19] that the duration of nuclear gamma emission process is about $\lambdabar/c$ ($\lambdabar$ is a wave length of gamma radiation divided by $2\pi$, $c$ – light velocity). For $^{109m}$Ag this duration is about $7.5\times10^{-21}$ sec that is much less than $t_{dd}$. If gamma emission (absorption) was completed such quickly then the large gamma line broadening would be inevitable  The nucleus may be considered in this case as an object in the constant magnetic field. Its gamma line would be split by this field, and so far as the fields on various nuclei are different the total picture of gamma line splitting becomes very broad. If the processes of gamma emission and absorption are prolonged then one can consider two following possibilities. The first one: the changing energy of dipole-dipole-interaction is being averaged to small value during the protracted nuclear radiation process. However there are serious theoretical objections against this hypothesis [20]. The second one: the nucleus and gamma radiation are not sensitive to the external influence until the radiation  process would be finished. The following reasoning may be considered as a support of this hypothesis. If a nucleus and gamma radiation can interact with environment during the radiation process then it is possible to register the gamma quantum by means of any detector before the end of its emission process. This act may happen in any moment with the same probability and on average – at the half of the possible duration of radiation process, that is at the half of mean life time of the nuclei in the excited state. This means that the minimum measurable width of a gamma line must be equal to the doubled natural width. However one observed in some Mössbauer experiments the gamma lines which widths were near the natural ones.

This work was performed under the support of RFBR (grant 09-02-00550a).